\documentclass{kapproc} % Computer Modern font calls
\usepackage{epsfig}

\newcommand{\Msun}{{M_\odot}}

\newcommand{\magn}{{^{25}\rm Mg}}
\newcommand{\alu}{{^{26}\rm Al}}
\newcommand{\iron}{{^{60}\rm Fe}}

\normallatexbib

\begin{document}

\articletitle[Radioactivies in Population Studies]{Radioactivities in Population Studies:\\
$\alu$ and $\iron$ from OB Associations}

\author{S.Pl\"uschke, R.Diehl, K.Kretschmer}
\affil{MPI f. extraterrestrische Physik, PO-Box 1312, D-85741 Garching, Germany}
\vspace{-0.5cm}
\author{D.H.Hartmann}
\affil{Department of Physics \& Astronomy, Clemson University, Clemson, SC 29634, USA}
\vspace{-0.5cm}
\author{U.Oberlack}
\affil{Astrophysics Laboratories, Columbia University, New York, NY 10027, USA}


\begin{abstract}
The observation of the interstellar 1.809 MeV decay-line of radioactive $\alu$ by the imaging gamma-ray 
telescope COMPTEL have let to the conclusion, that massive stars and their subsequent core-collapse 
supernovae are the dominant sources of the interstellar $\alu$ abundance. Massive stars are known to affect 
the surrounding interstellar medium by their energetic stellar winds and by the emission of ionising 
radiation. We present a population synthesis model allowing the correlated investigation of the gamma-ray 
emission characteristics with integrated matter, kinetic energy and extreme ultra-violet radiation emission of 
associations of massive stars. We study the time evolution of the various observables. In addition, we 
discuss systematic as well as statistical uncertainties affecting the model. Beside uncertainties in the input 
stellar physics such as stellar rotation, mass loss rates or internal mixing modifications due to a unknown 
binary component may lead to significant uncertainties.
\end{abstract}

\section{Introduction}
During recent years spatially resolved observations of decaying radioactive isotopes spread in the
interstellar medium of the Milky Way by their characteristic $\gamma$-ray emission became feasible.
The Compton telescope COMPTEL \cite{Sch93} aboard NASA's Compton Gamma-Ray Observatory surveyed 
the sky in the MeV regime during it's 9 year mission time. One of the key results of this mission
was the generation of the first all sky image in the 1.809 MeV $\gamma$-ray line from radioactive 
$\alu$ \cite{Obe96,Plu00a}. The COMPTEL observations confirm a integral $\alu$ content of the
Milky Way of 2 to $3\Msun$. In principle $\alu$ can be produced by various nucleosynthesis sites.
Since it's first detection by HEAO-C \cite{Mah84} non-explosive sites such as AGB stars \cite{Mow00}
and Wolf-Rayet stars \cite{Mey97} as well as explosive environments (novae \cite{Jos97} and
core-collapse supernovae \cite{WW95,WLW95}) are discussed as possible source candidates for
interstellar $\alu$.\\ 
The reconstructed $\gamma$-ray intensity distribution correlates best with tracers of massive stars 
such as thermal free-free emission from the ionised medium \cite{Kno99}. Due to its long lifetime 
freshly released $\alu$ travels quite some distances until its final decay. Therefore the 1.809 MeV
line emission is rather diffusive and it was not possible to observe isolated candidate sources. 
$\gamma^2$ Velorum, which is the nearest Wolf-Rayet star, is the only possible source for which
significant upper limits on the 1.8 MeV flux could be extracted so far \cite{Obe00}. Because of 
the close correlation between the galactic distribution of massive stars and the observed 1.8 MeV
intensity pattern, aggregations of massive stars such as young open clusters and OB associations
seem to be suitable laboratories to study these candidate sources. In particular, the by combined
analysis of the measured 1.8 MeV intensity together with observables in other wavelength bands one
may gain insight in the physical processes involved. Massive stars not only eject freshly synthesised
material into the surrounding ISM but also impart a huge amount of kinetic energy by stellar winds
and their subsequent core-collapse supernovae. Furthermore they affect the state of the interstellar
medium by emitting a large fraction of their electromagnetic radiation in the photoionizing extreme
ultra-violet regime. Therefore correlations between the $\gamma$-ray line emission of $\alu$ and
observables being related to one or the other phenomenon of massive stars are expected.\\
We make use of a detailed population synthesis model in combination with an 1-dimensional thin shell
expansion model of evolving superbubbles to study the time evolution of the different emission parameters 
and their resultant observables. In the following sections we present the model and discuss the predictions
and their uncertainties in detail. Finally, we apply the model to OB associations in the Cygnus region and
discuss the results for the expected $\alu$ content of this area.
\section{Modelling the OB star Emission}
\subsection{Release of radioactive Isotopes}
$\alu$ is synthesised via proton capture onto $\magn$ and therefore is a secondary product of the CNO-cycle
in massive stars. The efficient production of $\alu$ requires temperatures of at least $3.5\times10^7\,\rm K$,
which are reached by stars initially more massive than $25\Msun$. The production rate increases as the core 
temperature approaches values of $\sim10^8\,\rm K$ and depends strongly on the initial metallicity of the 
stellar material \cite{Mey97}. During later burning stages $\alu$ is produced by proton captures of 
secondary protons onto freshly synthesised $\magn$. The main destruction channels at temperatures below
220 keV are the $\beta^+$-decay and subsequently proceeding reactions with $\alu$ as input whereas for the
late burning phases the direct decay of the isomeric form of $\alu$ which is 220 keV above the ground-state 
destroys the remaining 26Al very efficiently. The internal mixing processes bring freshly synthesised $\alu$
into areas of the massive star which will be expelled by the strong stellar winds during the Of- and WR-phases.
Meynet et al. \cite{Mey97} computed a stellar evolution grid with detailed nucleosynthesis of $\alu$. Their
yields could be sufficiently well described by a power-law fit, which is used in our population synthesis
model to describe the $\alu$ release of Wolf-Rayet stars. The uncertainties in the description of the mixing
processes, stellar rotation and mass loss give raise to uncertainties in the expected yields of factors 
up to 3.\\
Beside Wolf-Rayet stars core-collapse supernovae are demonstrated to be efficient sources of interstellar $\alu$
\cite{WW95,WLW95,WH99}. In these events $\alu$ is produced by hydrostatic and explosive Ne-burning as well as
the $\nu$-process. Whereas for the mass range below $30\Msun$ detailed explosive nucleosynthesis models exist,
which cover a wide range of input physics including different mixing schemes and stellar rotation, for type
Ib/c supernovae this aspect is rather unexplored. The typical type II yields are fit reasonably well by a
power-law. In contrast, the convergence of the core-masses for the most massive stars lead to the assumption
of a non-mass-dependent 26Al yield, which is supported by the findings of Woosley et al. \cite{WLW95}. As in
the case of the WR contribution the theoretical uncertainties give raise to an inaccuracy of a factor 2 to 3.
In addition, core-collapse supernovae release a large amount of radioactive $\iron$, which migth be observed
by the $\gamma$-ray lines of its daughter nucleus $^{60}\rm Co$ at 1.173 and 1.332 MeV. Due to its solely
production by supernovae and the longer lifetime, which reduces the intensity of the expected $\gamma$-rays,
the Ironlines are still undetected so far. Nevertheless, the ejection of $\iron$ is included in our population
synthesis model.\\
\begin{figure}[h!]
 \vspace{-0.75cm}
 \begin{minipage}[h!]{8cm}
  \epsfig{figure=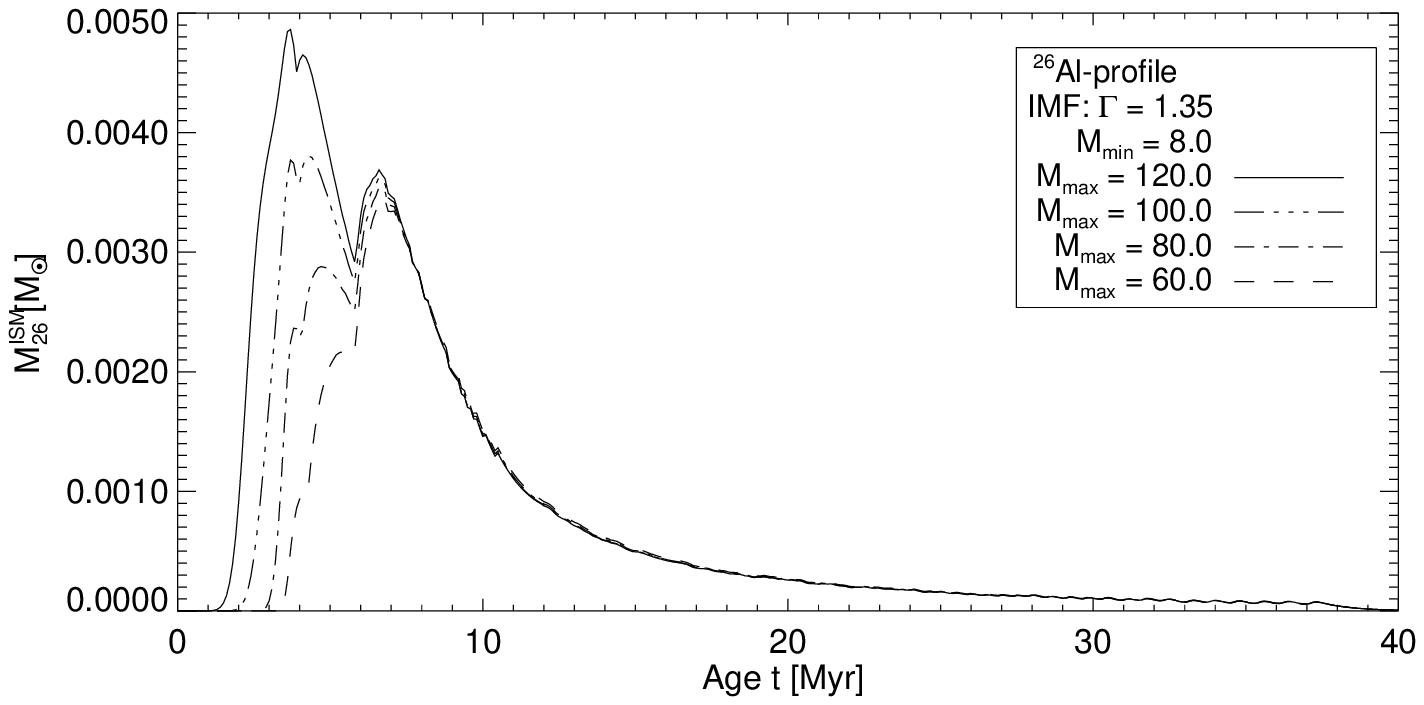,width=7.75cm}
 \end{minipage}
 \begin{minipage}[h!]{3.5cm}
  \narrowcaption{Time evolution of the interstellar mass of $\alu$. The curves assume an OB association of
   500 stars initially more massive than $8\Msun$ with a paramterised upper mass limit ditribute as a
   Salpeter-IMF (see legend).}
   \vspace{0.6cm}
  \label{fig1}
 \end{minipage}
 \vspace{-0.75cm}
\end{figure}
Figure \ref{fig1} shows the time evolution of interstellar content of $\alu$ for an OB association assuming
an instantaneous starburst with Salpeter-IMF. The resulting time profile is convolution of the mass spectrum
and the mass- and time-dependent ejection rates of $\alu$ taking the radioactive decay into account. The profile 
shows the typical two-peaked structure, where the first peak is due to the WR contribution and the second originates
from the SN activity. Usually OB associations have limited population statistics, leaving the observer with a 
poor sampling of the initial mass spectrum. A Monte Carlo version of our population synthesis code allows the
estimation of the uncertainties introduced by this effect. Our finding is, that for populations being initially 
richer than 100 stars more massive than $8\Msun$, the resulting time profiles are dominated by theoretical
uncertainties instead of statistical effects.\\
An additional source of uncertainty is a possible contribution to the interstellar $\alu$ content by peculiar
massive close binaries. Langer et al. \cite{La98} have shown that some of these systems may give raise to a large 
overproduction of $\alu$ in the supernova explosion of secondary. Their contribution may be enhanced by factors
up to 1000. Up to now, it is not possible to really incorporate these effects in a population synthesis model 
due to a lack of a detailed investigation of the appropriate parameter space. Therefore we have to leave this as
open question.
\subsection{Imparting kinetic Energy to the ISM}
\begin{figure}[h!]
 \vspace{-0.75cm}
 \begin{minipage}[h!]{8cm}
  \epsfig{figure=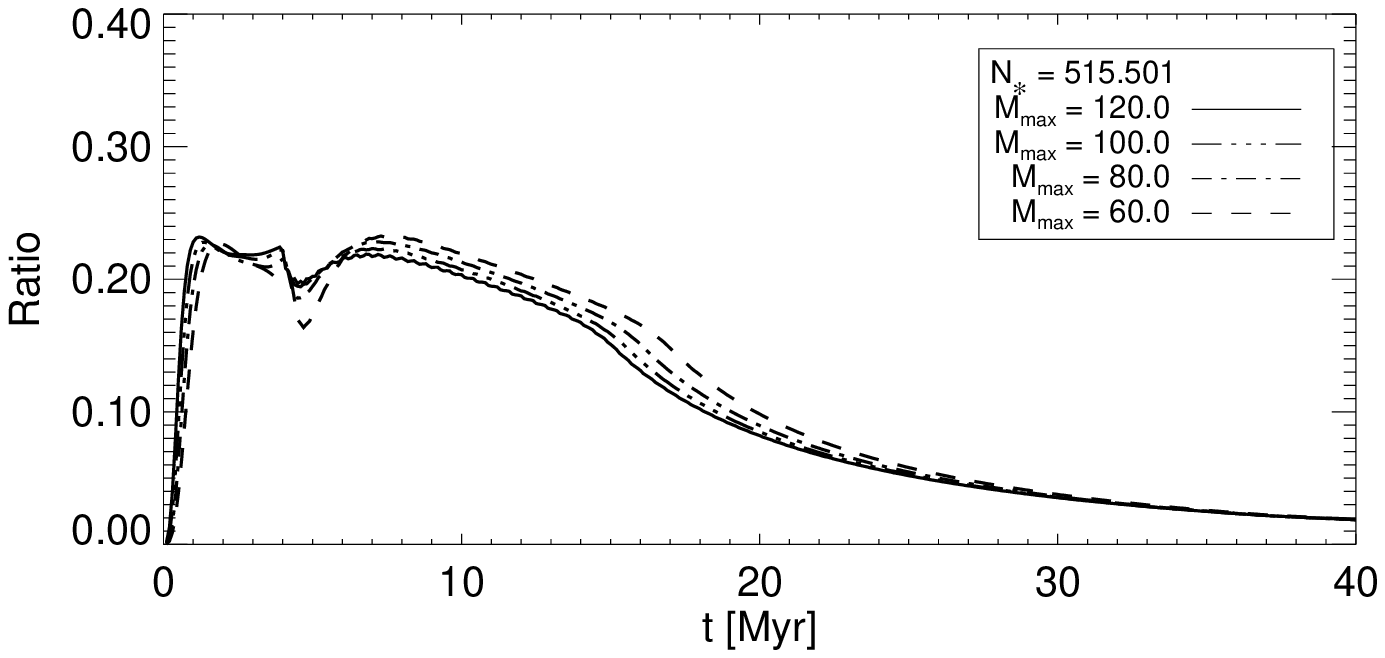,width=7.75cm}
 \end{minipage}
 \begin{minipage}[h!]{3.5cm}
  \narrowcaption{Ratio of observable to injected kinetic energy of an expanding superbubble around an OB association.
   The surrounding density was chosen to 100 cm-3 whereas the evaporation parameters were set to $\rm\tau_{poros}
   =0.005$ and $\rm\tau_{evap} = 0.5\, Myr$.}
   \vspace{0.6cm}
  \label{fig2}
 \end{minipage}
 \vspace{-0.75cm}
\end{figure}

A well known phenomenon of massive stars is their very strong stellar wind imparting a fairly large amount of
kinetic energy to the ISM. In addition, their subsequent supernova explosions release on a very short time-scale 
even larger amounts of energy to the ISM. In OB associations these effects can be expected to overlap and serve
as an energy source for blowing large-scale gas structures, known as superbubbles, into the surrounding ISM. Our 
population synthesis model predicts the flux of kinetic energy as well as matter due to an underlying massive
star population. The stellar wind part is modelled using the semi-empirical mass loss rates from the Geneva stellar 
evolution models \cite{Mey94} in combination with a semi-analytical wind velocity formula from Prinja and colleagues
\cite{HP89,PBH90}. The energy released by a single supernova is assumed to be $\rm10^{51}\, erg$ per event. By using
a similar scheme as for the nucleosynthesis part our population synthesis model predicts the integral flux of 
energy and matter from an association.\\
The integral wind power as well as the flux of matter can only be observed by their impact on the surrounding ISM,
therefore we used the results from our population synthesis model as input for an 1-dimensional, numerical model of 
an expanding supershell. The model is based on the thin shell approximation and incorporates radiative cooling as
well as a parameterised description of evaporation from the inner shell or cloudlets which passed the shock and made
it into the hot bubble medium. Our description of a possible evaporative poisoning as it was labelled by Shull \&
Saken \cite{SS95} aims on a study of the relevance of this process and uses two parameters, which are the transmission 
efficiency $\rm\epsilon_{poros}$ and the evaporation time-scale $\rm\tau_{evap}$. Depending on the ambient density
and pressure evaporation becomes critical for quite low transmission efficiencies of the order of few thousands
relative to the swept up mass sitting in the thin shell. This is especially true if the evaporation time-scale is
short. If the bubble medium reaches a critical density at sufficiently low temperatures of some $\rm10^6\, K$ the
interior energy of the bubble is converted into radiation and the bubble medium is cold down to some $\rm10^4\, K$
rather immediately. If the evaporation time-scale is long compared with the dynamic time-scales then the transmission
of some fractions of the ambient medium reduces the mass of the shell at a given time and the bubble expansion stays
faster than in cases where no material is transmitted through the shell. In this cases the bubbles become larger.
Figure \ref{fig2} shows the time evolution of the ratio of the observable kinetic energy of an expanding supershell
relative to the kinetic energy from source. In this case intermediate values have been chosen for the evaporation 
parameters.
\\
During the early phase the ratio approaches the canonical value of 20-25\% rather quickly, whereas after
approximately 15 Myr the ratio drops significantly and approaches values of 5\% or less in the late phase.
This behaviour depends strongly on the chosen parameter set for the environment as well as evaporation 
process. If evaporation is efficiently suppressed due to magnetic fields or some other processes the ratio
stays near 20\% over the whole evolution. But, the critical parameter range for efficient cooling and
subsequently stalling the expansion are limited by very modest values, which might be typical for interstellar 
conditions.\\
The combination of these gas dynamic effects with the emission characteristics for the release of freshly produced
radioactive isotopes such as $\alu$ and $\iron$ allows to predict the spatial intensity distribution of the expected
$\gamma$-ray emission. This is of special importance in the physical understanding why the observed 1.809 MeV
intensity distribution correlates so well with strongly blurred tracers. Beside the uncertainties due to the 
evaporation process and therefore the efficiency of radiative cooling in the early stages, the results of the
population synthesis are strongly affected by the uncertainties in the stellar wind models. The Geneva stellar
evolution which have been used for consistency reasons apply an enhancement factor of 2 to the mass loss 
rates during the main-sequence and WR-phase. Therefore the expected wind power is roughly a factor of 2 higher
than for typical observed stars. In contrast, an underestimation of the wind velocities of only 30\% will already
restore the wind power in our model. We therefore estimate our results to be reliable within a factor of 2 or so.
\subsection{Ionising Radiation}
As mentioned in the introduction the observed 1.809 MeV intensity pattern correlates very nicely with the
galactic free-free emission observed with COBE DMR at 53 GHz \cite{Kno99,Ben94}. The interpretation of this
close correlation points to massive, hot stars being the dominant sources of interstellar $\alu$. The free-free 
emission originates from the ionised portions of the interstellar medium which is found in compact HII regions as
well as in from of the diffuse ionised medium. It is strongly believed that in both cases the ionising extreme
ultra-violet radiation from young, massive stars is the dominant source of keeping the medium ionised \cite{DS00}.
In the framework of our population synthesis model we therefore incorporated the emission of Lyman continuum photons
by means of a simple fit-function in dependence of the initial mass of the emitting star.
\begin{equation}
Q_{0/1}=\left[\exp\left(a_1+\frac{a_2}{M_i}\right)\right]\cdot10^{49}\,{\rm s^{-1}}
\end{equation}
This function is fitted to the Lyman continuum photon fluxes from detailed stellar atmosphere calculations
\cite{Vac96}. For main-sequence stars the error is less then 5\% by using this fit instead of the appropriate
model values for the given initial mass. Nevertheless, the error of the resulting time-profile of the Lyman 
continuum emission of an OB association is considerably larger due to disregarding the effects of stellar
evolution. However, a comparison with an alternative EUV emission model based on direct interpolation of the 
appropriate fluxes from stellar atmosphere models reveals an overestimation of the cumulative Lyc flux in our
association model of less than 30\% which is of the order as theoretical uncertainties of the underlying 
stellar models.
\begin{figure}[h!]
 \vspace{-0.25cm}
 \begin{minipage}[h!]{8cm}
  \epsfig{figure=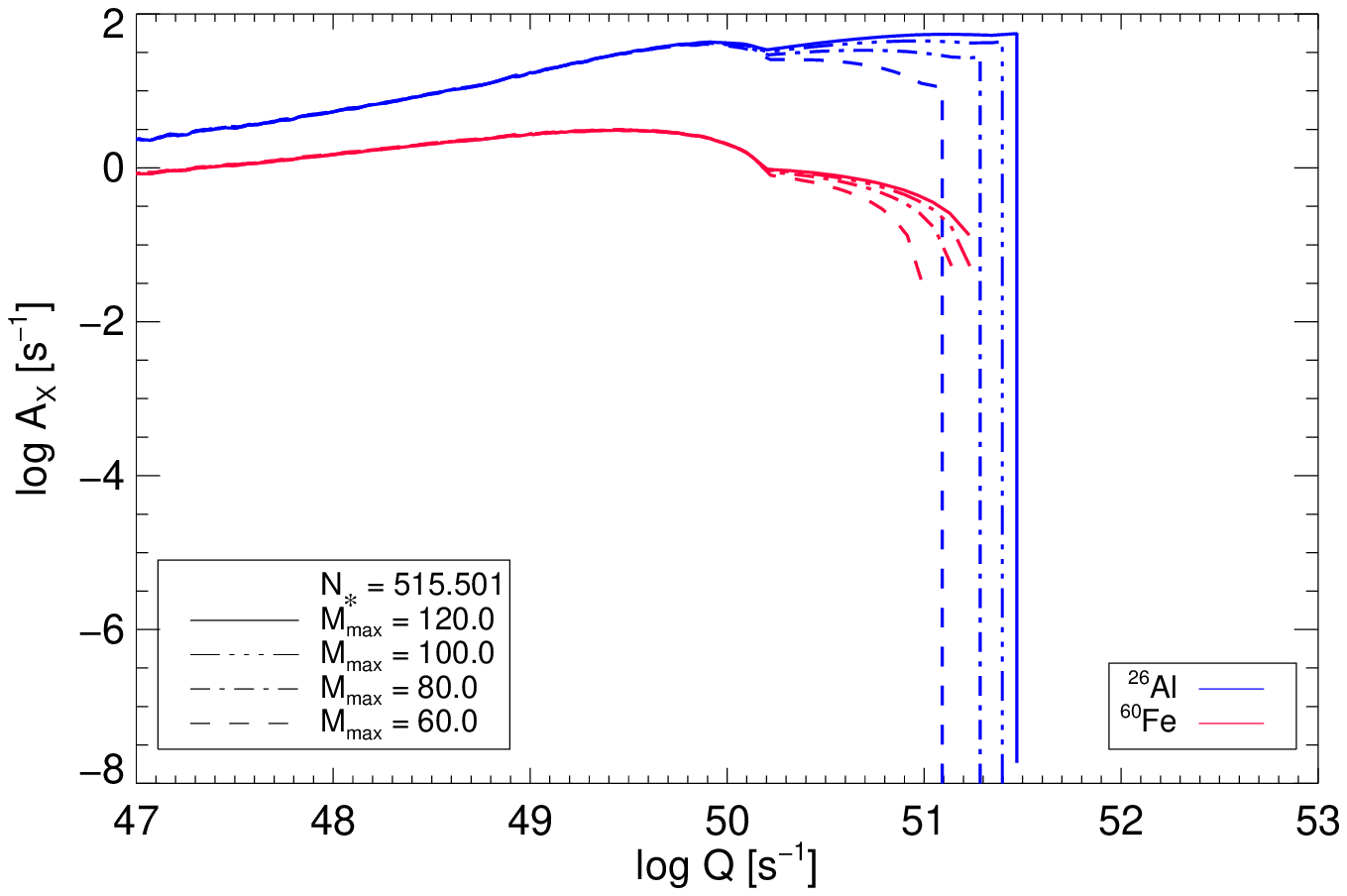,width=7.75cm}
 \end{minipage}
 \begin{minipage}[h!]{3.5cm}
  \narrowcaption{Evolution of the correlation of the Lyman continuum flux vs. the decay rate of interstellar 
   $\alu$ and $\iron$, respectively}
   \vspace{0.6cm}
  \label{fig3}
 \end{minipage}
 \vspace{-0.5cm}
\end{figure}\\
Figure \ref{fig3} shows the time-evolution of the integral Lyc flux versus the decay rate of interstellar
$\alu$ and $\iron$, respectively, for the same association model discussed earlier. The plotted trajectories
start with low decay rates at high EUV luminosities and evolve to considerable decay rates at vanishing Lyc 
fluxes. In principle, this behaviour could be exploit to construct a very sensitive age indicator for the
underlying population. The ratio of the observed 1.809 MeV flux from an astronomical source population and the 
respective Lyc flux, which in principle could be extracted from the observed free-free intensity by using a
proper ionisation model, should be independent of the distance to the population. After an additional
normalisation of this ratio relative to the Lyc flux of an O7V star one gets the O7V equivalent yield for
an radioactive isotope for a given population \cite{Kno97}. Indeed, this quantity can be used as sensitive
age indicator as shown in figure \ref{fig4} for the case of $\alu$.
\begin{figure}[h!]
 \vspace{-0.25cm}
 \begin{minipage}[h!]{8cm}
  \epsfig{figure=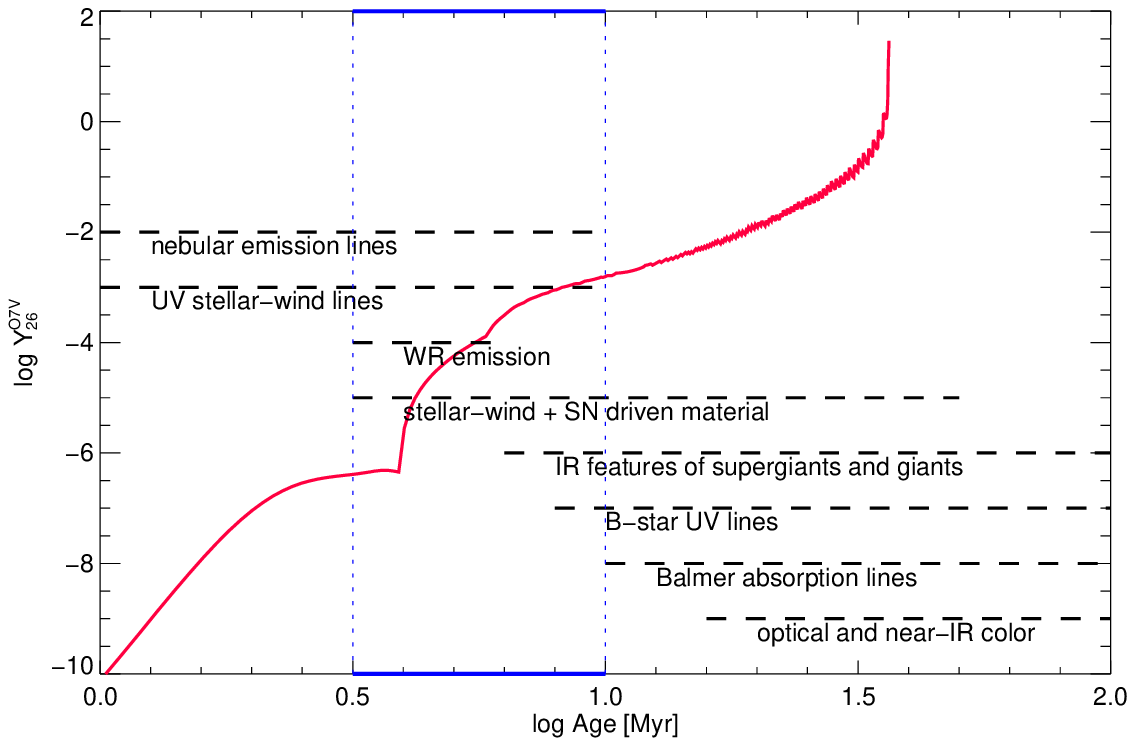,width=7.75cm}
 \end{minipage}
 \begin{minipage}[h!]{3.5cm}
  \narrowcaption{Time evolution of the O7V equivalent yield for $\alu$ in comparison to the age ranges for
   other age indicators in astronomical populations.}
   \vspace{0.6cm}
  \label{fig4}
 \end{minipage}
 \vspace{-0.25cm}
\end{figure}\\
Especially in the regime between 3 and 10 Myr the O7V equivalent yield shows a strong increase over 3 orders
of magnitude. This is due to the strong decrease of the number of O stars because of supernovae and the two 
emission peaks in the time profile of the interstellar $\alu$ content (cf. Fig. \ref{fig1}).\\
Additionally, the released Lyc flux can be used as input to an ionisation model to calculate a prediction map
of expected free-free intensity of an area under investigation. This could in turn be compared to the observed
one and therefore may give an additional constrain on the real population.
\section{Statistics}
\begin{figure}[h!]
 \vspace{-0.25cm}
 \begin{minipage}[h!]{8cm}
  \epsfig{figure=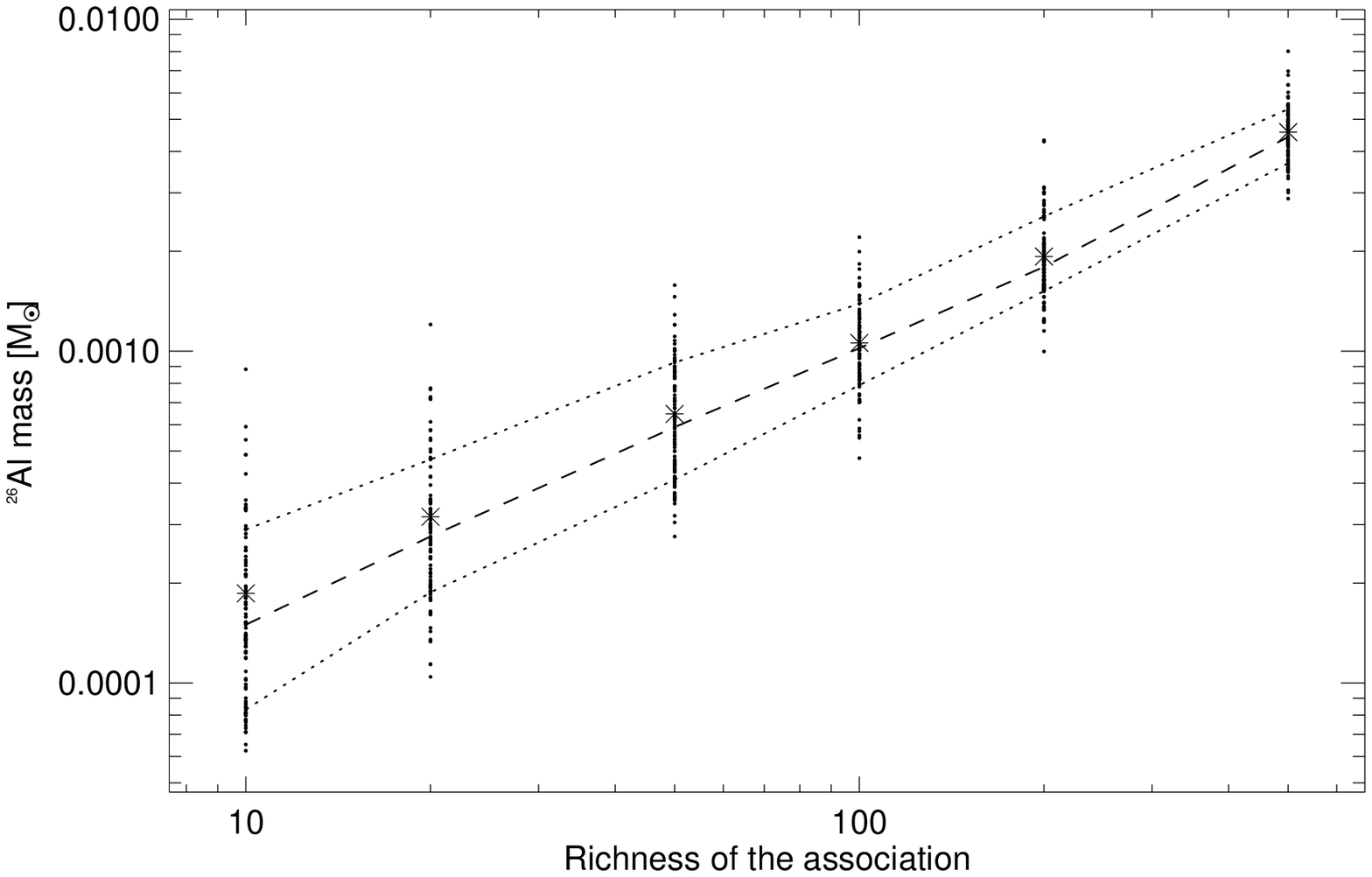,width=7.75cm}
 \end{minipage}
 \begin{minipage}[h!]{3.5cm}
  \narrowcaption{Statistical error of the interstellar $\alu$ mass as function of the richness of the
   association}
   \vspace{0.6cm}
  \label{fig5}
 \end{minipage}
 \vspace{-0.25cm}
\end{figure}

As already discussed in section 2.1 the application of our population synthesis model, which assumes a continuos
mass function, to a real astronomical population is expected to be significantly disturbed by population statistics. 
Whereas for galaxies the mass function is sampled sufficiently dense to justify a quasi-analytic treatment this
might not be the case for OB associations or even worth open clusters. We therefore studied the statistical errors as 
function of the richness by means of Monte Carlo version of the model. Figure \ref{fig5} shows the spread of the
resulting interstellar $\alu$ masses at time of the emission maximum in dependence of the richness.
\\
For populations richer than 100 stars initially more massive than $8\Msun$ the statistical error drops below
the theoretical uncertainties. In addition, the Monte Carlo code allows the determination of probability
density functions for direct application of the association model to observed populations by means of a Bayesian 
analysis.
\section{1.8 MeV from Cygnus}
\begin{figure}[h!]
  \begin{minipage}[h!]{5.5cm}
    \epsfig{figure=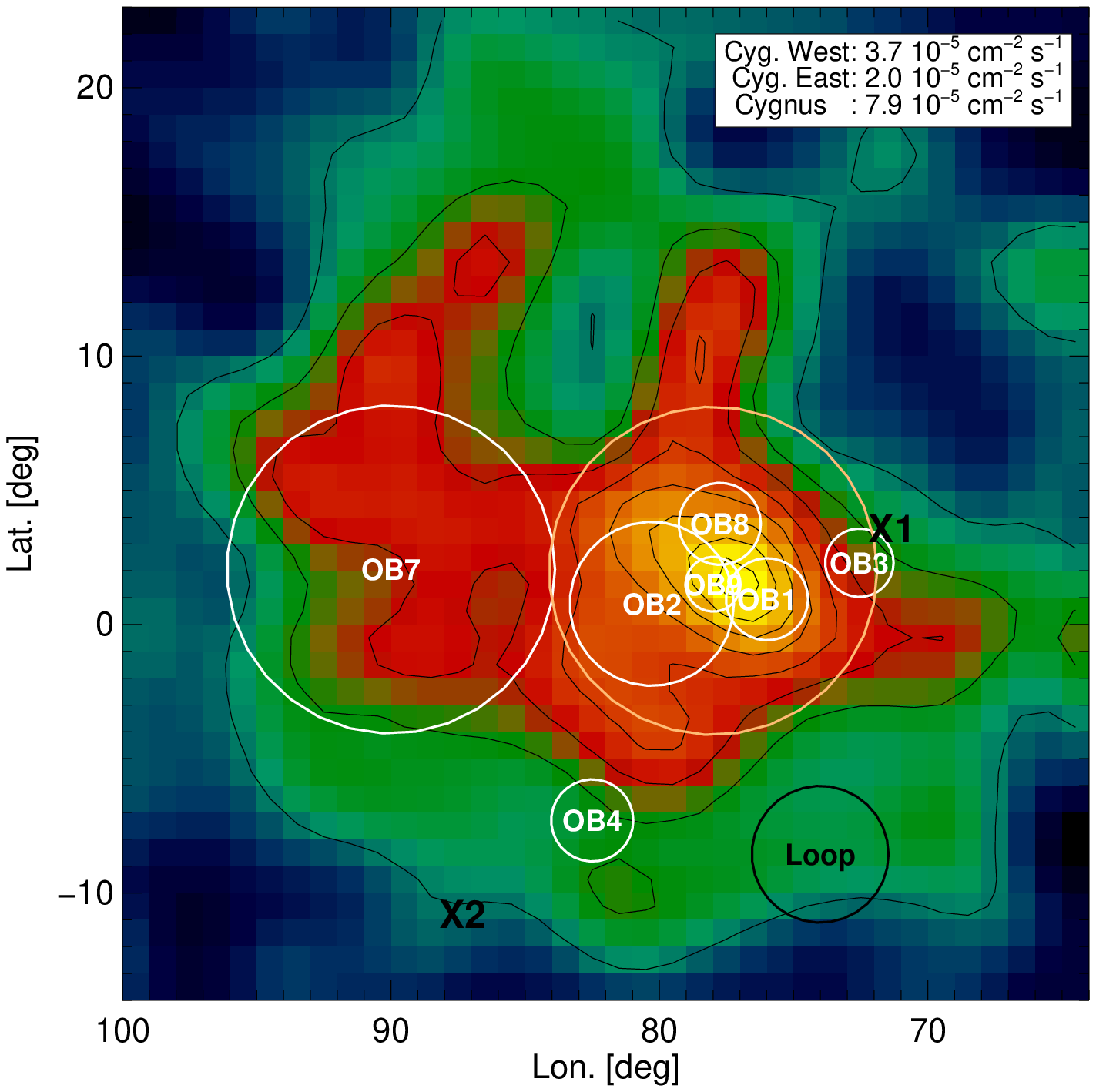,height=5cm}
    \caption{1.809 MeV emission from the Cygnus Region as observed with COMPTEL \cite{Plu00b}}
    \label{fig6}
  \end{minipage}
  \begin{minipage}[h!]{5.5cm}
    \epsfig{figure=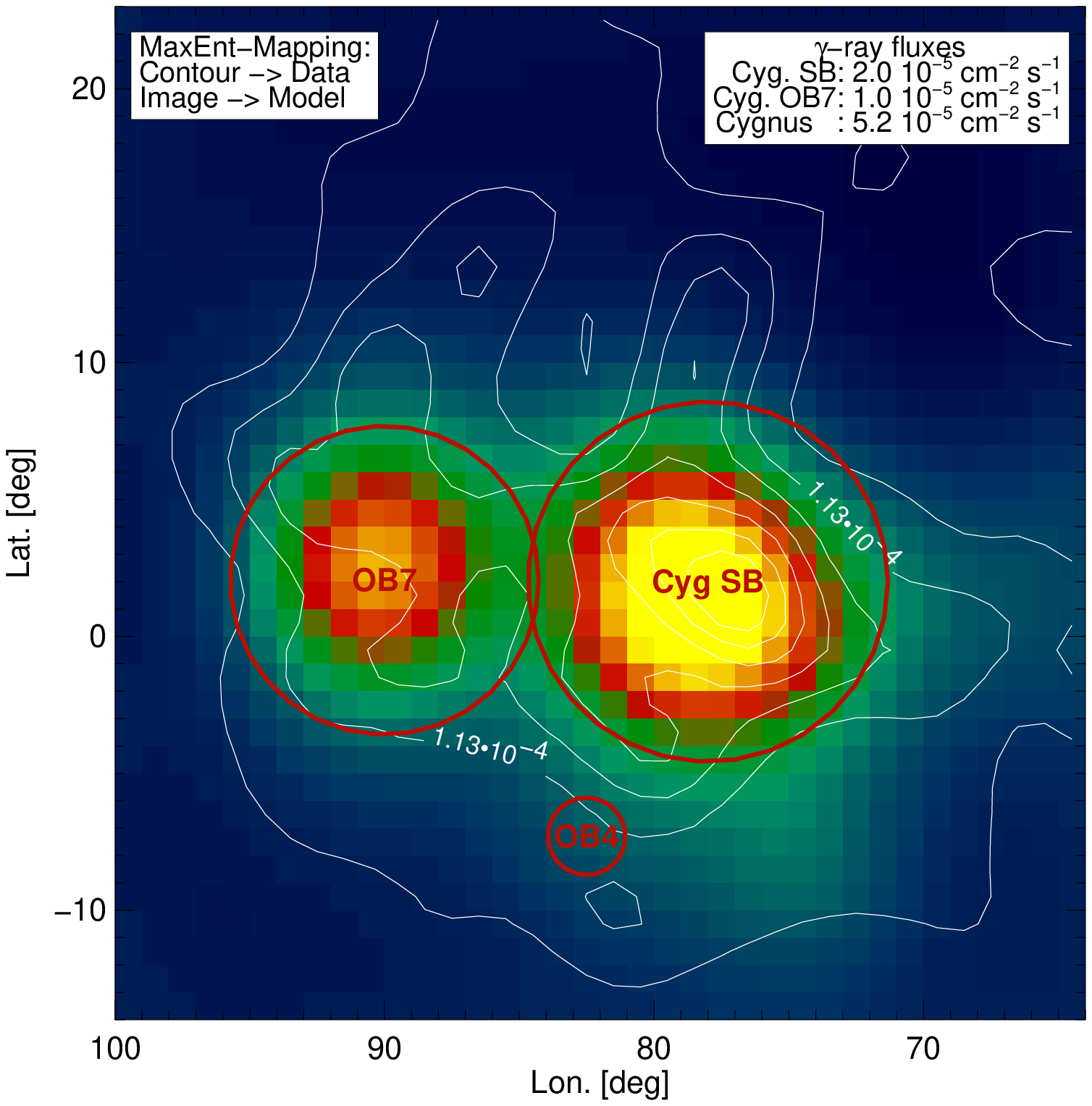,height=4.95cm}
    \caption{1.809 MeV intensity model based on population synthesis of the Cygnus OB associations}
    \label{fig7}
  \end{minipage}
\end{figure}

The Cygnus region is the most significant isolated structure in the COMPTEL 1.8 MeV maps beyond the inner galaxy.
Figure \ref{fig6} shows the Cygnus OB associations superimposed as circles on the latest COMPTEL Maximum 
Entropy image of the Cygnus region \cite{Plu00b}.\\
\\
As a first test we applied our model to the given OB associations encircled in Figure \ref{fig6} and 
generated an 1.809 MeV intensity model, which for simplicity neglects the gas dynamic effects 
discussed in section 2.2. The population synthesis is based on a database built from recent literature
on population studies of the Cygnus region. Figure \ref{fig7} shows the resulting intensity model.
At first glance this model already reproduces the observed emission peak and the extension to higher
longitudes. But, significant discrepancies are still remaining. First, the overall flux is underestimated
by the model. After a variation of the slope of the initial mass spectrum turned out to be insignificant in
resolving this problem, the most plausible explanation is an underestimation of populations due to strong 
visual extinction towards Cygnus. Second, the modelled intensity distribution is significantly narrower than
the observed pattern, which could be understood if one reminds the neglect of gas dynamic effects due to bubble
formation.
\section{Summary}
We have presented an extension of population synthesis studies to the domain of $\gamma$-ray line astronomy.
The discussed theoretical uncertainties are far from being insignificant. In addition, it is questionable if the 
observation and modelling of OB associations as laboratories for testing specific aspects of the theoretical
understanding of the chemical evolution leading to interstellar enrichment of radioactive isotopes are sufficient
due to possible tremendous statistical uncertainties. For stellar populations of massive stars being richer than
100 objects our model shows that theoretical uncertainties begin to dominate. So, one can conclude that indeed
areas with rich associations may be used for testing theoretical aspects. At least, the observations can be used 
to check the consistency by extracting constrains from combined analyses at different wavelengths.\\
It was shown explicitly that the combined analysis of free-free emission and the observed 1.809 MeV intensity may
serve as a sensitive age indicator for young stellar populations. Furthermore, the rather simplistic application of
the model to the 1.809 MeV emission in the Cygnus region already showed the potential in extracting constrains
from the $\gamma$-ray line observations. However, due to the neglect of additional possible source candidates such
as peculiar massive close binary systems these constrains are not very stringent up to now.
\bibliographystyle{apalike}

\begin{chapthebibliography}{}
\bibitem{Sch93}
Sch\"onfelder, V., et al.: 1993, ApJS 86, 657+

\bibitem{Obe96}
Oberlack, U., et al.: 1996, A\&AS, 120, 311-314

\bibitem{Plu00a}
Pl\"uschke, S., et al.: 2000, accepted for publ. in 'Proc. of the $4^{\rm th}$
INTEGRAL Workshop', Alicante/Spain, publ. by ESA

\bibitem{Mah84}
Mahoney, W.A., et al.: 1984, ApJ, 286, 578-585

\bibitem{Mow00}
Mowlavi, N. \& Meynet, G.: 2000, A\&A, 361, 959-976

\bibitem{Mey97}
Meynet, G., et al.: 1997, A\&A, 320, 460-468

\bibitem{Jos97}
Jos\'e, J., et al.:1997, ApJ, 479, L55+

\bibitem{WW95}
Woosley, S. \& Weaver, T.: 1995, ApJS, 101, 181+

\bibitem{WLW95}
Woosley, S., et al.: 1995, ApJ, 448, 315+

\bibitem{WH99}
Woosley, S. \& Heger, A.: 1999, Proc. of 'Astronomy with Radioactivities II",
edt. by R.Diehl \& D.Hartmann, MPE Rep. 274, 133-140

\bibitem{Kno99}
Kn\"odlseder, J., et al.: 1999, A\&A, 344, 68-82

\bibitem{Obe00}
Oberlack, U., et al.: 2000, A\&A, 353, 715-721

\bibitem{La98}
Langer, N., et al.: 1998, Proc. of the $9^{\rm th}$ Workshop on Nuclear
Astrophysics, edt. by W.Hillebrandt \& E.M\"uller, MPA Garching, p. 18+

\bibitem{Mey94}
Meynet, G., et al.: 1994, A\&AS, 103, 97-105

\bibitem{HP89}
Horwarth, I. \& Prinja, R.: 1989, ApJS, 69, 527-592

\bibitem{PBH90}
Prinja, R., et al.: 1990, ApJ, 361, 607-620

\bibitem{SS95}
Shull, J.M. \& Saken, J.: 1995, ApJ, 444, 663-671

\bibitem{Ben94}
Bennett, C.L., et al.: 1994, ApJL, 464, L1+

\bibitem{DS00}
Dove, \& Shull, J.M.: 2000, ApJ, 531, 846-860

\bibitem{Vac96}
Vacca, W., et al.: 1996, ApJ 460, 914+

\bibitem{Kno97}
Kn\"odlseder, J.: 1997, PhD Thesis Univ. Toulouse

\bibitem{Plu00b}
Pl\"uschke, S., et al.: 2000, accepted for publ. in 'Proc. of the $4^{\rm th}$
INTEGRAL Workshop', Alicante/Spain, publ. by ESA

\end{chapthebibliography}

\end{document}